\documentclass[%
 aip,
 amsmath,amssymb,
 reprint,%
floats
]{revtex4-1}

\usepackage{graphicx}
\usepackage{dcolumn}
\usepackage{bm}
\usepackage[dvipsnames]{xcolor} 
\usepackage[french, main=english]{babel}

\usepackage[utf8]{inputenc}
\usepackage[T1]{fontenc}
\usepackage{mathptmx}
\usepackage{etoolbox}
\usepackage{hyperref}

\makeatletter
\def\@email#1#2{%
 \endgroup
 \patchcmd{\titleblock@produce}
	{\frontmatter@RRAPformat}
	{\frontmatter@RRAPformat{\produce@RRAP{*#1\href{mailto:#2}{#2}}}\frontmatter@RRAPformat}
	{}{}
}%
\makeatother

\newcommand{\unit}[1]{\;\mathrm{#1}}

\begin{document}
	
\preprint{AIP/123-QED}

\title[Unified non-equilibrium simulation methodology for flow through nanoporous carbon membrane]{Unified non-equilibrium simulation methodology for flow through nanoporous carbon membrane}

\author{Geoffrey Monet}

\author{Marie-Laure Bocquet}
\author{Lydéric Bocquet}
	\email{lyderic.bocquet@phys.ens.fr}

\affiliation{Laboratoire de Physique de l’École Normale Supérieure, ENS CNRS, Paris, France}%

\date{\today}

\begin{abstract}
	The emergence of new nanoporous materials, based {\it e.g.} on 2D materials, offers new avenues for water filtration and energy. There is accordingly a need to investigate the molecular mechanisms at the root of the advanced performances of these systems in terms of nanofluidic and ionic transport. In this work, we introduce a novel unified methodology for Non-Equilibrium classical Molecular Dynamic simulations (NEMD), allowing to apply likewise pressure, chemical potential and voltage drops across nanoporous membranes and quantifying the resulting observables characterizing confined liquid transport under such external stimuli. We apply the NEMD methodology to study a new type of synthetic Carbon NanoMembranes (CNM), which have recently shown outstanding performances for desalination, keeping high water permeability while maintaining full salt rejection. The high water permeance of CNM, as measured experimentally, is shown to originate in prominent entrance effects associated with negligible friction inside the nanopore. Beyond, our methodology allows to fully calculate the symmetric transport matrix and the cross-phenomena such as electro-osmosis, diffusio-osmosis, streaming currents, etc. In particular, we predict a large diffusio-osmotic current across the CNM pore under concentration gradient, despite the absence of surface charges. This suggests that  CNMs are outstanding candidates as alternative, scalable membranes for osmotic energy harvesting.

\end{abstract}

\maketitle

\begin{quotation}

\end{quotation}

\section{Introduction}
	The emergence of nanomaterials, such as carbon nanotube, graphene, and 2D materials in general, has triggered much interest in the context of their use as a membrane for water filtration \cite{park_freeman_2017}. In this context, carbon materials were systematically found to outperform other materials in terms of water permeability or separation efficiency \cite{faucher_strano_2019, bocquet_bocquet_2020}. This puzzling result has triggered many fundamental investigations, with the emergence of nanofluidics, the science exploring the molecular mechanics of nanometer flows. In particular, membranes made of carbon nanotubes (CNT) were shown to exhibit huge water permeabilities down to (sub-) nanometer porosities \cite{holt_bakajin_2006, secchi_bocquet_2016, tunuguntla_noy_2017}. This puzzling performance has been rationalized theoretically by invoking the radius-dependent role of quantum excitations between collective modes of water and plasmons in the multiwall CNT in the terahertz regime\cite{kavokine_bocquet_2022}.\par
	
	Apart from CNTs, carbon nanomaterials in various forms have been considered for water filtration \cite{wang_karnik_2017}: membranes made of nanotube porins \cite{holt_bakajin_2006,tunuguntla_noy_2017}, graphene membranes \cite{celebi_park_2014, wang_karnik_2017}, graphene oxide membranes \cite{abraham_nair_2017} and nanoporous graphene \cite{cheng_kidambi_2022}. These systems exhibit outstanding performances in the context of filtration, but a formidable challenge with these nanomaterials though, is the scale-up process, mandatory for practical applications.\par

	In this context, the recently synthesized carbon nanomembranes (CNM) are promising \cite{yang_golzhauser_2018}: CNMs are ultra-thin, but large-scale, membranes made of carbon material, with high densities of sub-nanometric porosities. They exhibit excellent ion separation and very high water permeabilities \cite{yang_golzhauser_2018}. Quantitatively, the nanopore exhibit a size in the range of a few Angströms, with high densities, $\sim 10^{13}-10^{14} \unit{pores/cm^2}$, and CNM permeabilities reach values $\sim 3000-7000 \unit{L/m^2/h/MPa}$, which is several orders of magnitude higher than typical reverse osmosis membranes ($\sim$ a few $\unit{L/m^2/h/MPa}$). 
	Now the results for the outstanding permeability of CNM, together with an excellent selectivity, raises the question of its underlying mechanism. This permeability/selectivity balance is usually a trade-off \cite{park_freeman_2017}, which CNM membranes seem to bypass to some extent.\par

	To date, the CNM membrane lacks a fundamental understanding of the fluid transport mechanism. To this end, we aim at proposing a realistic atomistic model for CNM and developing classical non-equilibrium molecular dynamics simulations suitable to all membrane systems in general, to rationalize the transport of nanoconfined water and salt through it.

	In terms of computational methodology, one general difficulty of non-equilibrium simulations is to apply thermodynamic driving forces across a non-translationally invariant system, here nanopores pierced in a membrane, while using globally periodic boundary conditions. This is specifically important for transport across nanoporous membranes where entrance effects are expected to play a central role in the transport. For pressure-induced water flow across nanopores, an approach proposed by Goldsmith and Martens consists in applying external forces to selected atoms in the liquid water reservoir and computing the induced pressure drop \cite{goldsmith_martens_2009}. We show that the approach can be generalized to any thermodynamic force in order to generate pressure, chemically and electrically-induced flows across a nanopore. It makes use of various sets of external forces applied on a slab of atoms inside the reservoir and inferring the corresponding pressure, concentration and potential drops associated with the specific external forces. This non-equilibrium methodology allows us to quantify pressure-, concentration- and voltage-driven transport across the CNM, and derive the complete transport matrix \cite{yoshida_barrat_2014a}. We can then compare the various transport mobilities to analytical predictions based on continuum frameworks. This allows to disentangle the mechanisms at the core of the underlying transport phenomena, and specifically address the limiting role of access effects. Overall our results highlight the geometric specifications that govern the reported performances of CNM membranes.\par

\section{Regular atomistic model for CNM}
\begin{figure}[t!] 
		\begin{small}
			\begin{center}
				\includegraphics{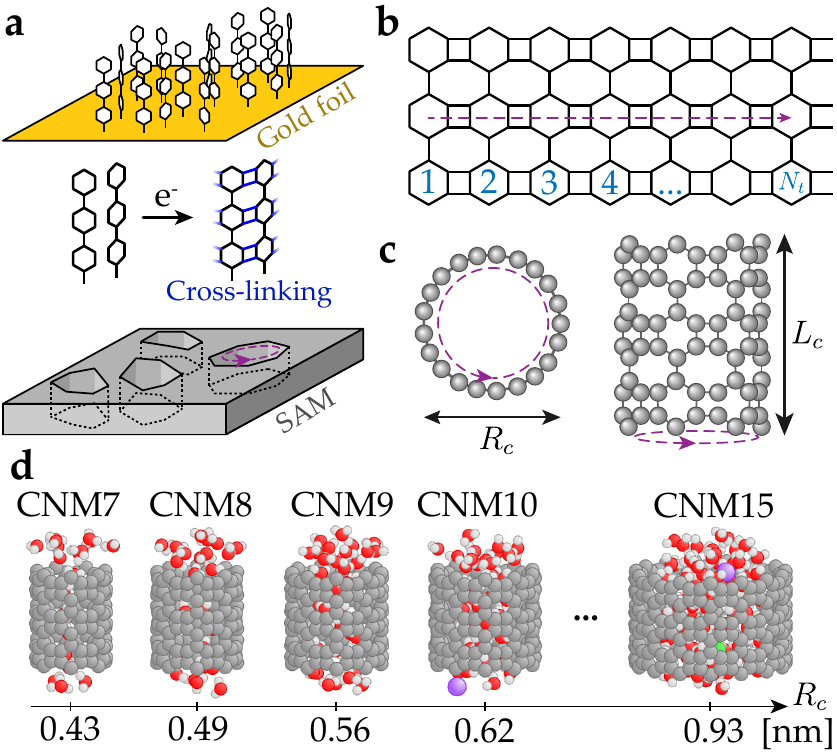}
			\end{center}
			\caption{
				(a) Illustration of the sequential synthesis of CNM.
				(top) Close-packed grafting of TPT molecules via sulfur heteroatom onto a gold surface. 
				(middle) Electron irradiation upon TPT molecules yielding cross-links between them. 
				(bottom) After substrate removal, resulting array of roughly circular subnanometer pores. 
				(b) Construction of the atomistic regular pore model : planar structure formed by perfect C-C coupling between $N_t$ TPT molecules. A regular nanopore model is formed from the wrapping of this planar structure along the dashed purple vector.
				(c) Top and side views of a regular CNM nanopore for $N_t=7$ labelled CNM7 with its two geometric specifications $R_c$ and $L_c$
				(d) Structures of studied nanopores with various radii $R_c$ in $\unit{nm}$ indicated in the rule below and their corresponding labels. The CNMs are filled with water and NaCl salt ($Na^+$ in green and $Cl^-$ anion in purple)}.
			\label{fig:nanopore}
		\end{small}
	\end{figure}

	The CNMs introduced in Ref.\cite{yang_golzhauser_2018} are nanoporous carbon-based membranes with high densities of sub-nanometric pores and nanometric thickness. Their synthesis procedure is depicted in figure~\ref{fig:nanopore}a. TPT (TerPhenylThiol) molecules are covalently grafted on a gold substrate via its ending sulfur atom. As such, the TPT molecules form a close-packed array. The adsorbed molecules are then exposed to an electron beam which leads to a dehydrogenation process and the formation of novel C-C bonds between the adjacent adsorbed TPT (see new C-C bonds represented in blue on the figure~\ref{fig:nanopore}a). Eventually, the resulting carbon overlayer is separated from the original gold substrate and the anchoring sulfur atoms. The as-obtained two-dimensional nanoporous structure is thus a compact self-assembled monolayer (SAM) only made of cross-linked precursor molecules. It leads to an array of non-regular nanopores of subnanometer sizes and of fixed thickness of $1 \unit{nm}$(see Figure~\ref{fig:nanopore}a bottom).\par

	In this study, we model the closed pores of CNM with a tubular model obtained after the rolling up of a layer formed of $N_t$ fully cross-coupled TPT molecules (see figure~\ref{fig:nanopore}b). Using this regular model, it is straightforward to tune its geometry by changing its radius $R_c$ and to quantify such effect on the fluid transport mechanisms at play. 
	In previous experimental studies \cite{yang_golzhauser_2018}, transmission electron microscopy pictures revealed a pore diameter distribution that ranges around $0.7\unit{nm}$. In our regular nanopore model, this average diameter is reached with the wrapping of 7 TPT molecules - CNM7 - as represented in figure~\ref{fig:nanopore}c. The length of the nanopore is fixed by the length of a TPT molecule, i.e. $1 \unit{nm}$. Hence in CNM pores the aspect ratio - radius $R_c$ versus thickness $L_c$ - is close to 1. 
	Interestingly this carbon system lies in between nanoporous graphene membranes and carbon nanotubes for which the fluid transport properties strongly differ. For the former, entrance effects dominate the transport \cite{zhao_yan_2014}, while for the latter the micrometer length induces a key role of the internal friction although the water friction is unexpectively low for nanometric carbon nanotubes \cite{secchi_bocquet_2016a}. This following study will decipher the type of transport mechanism through CNM with a specific focus on the diverse role of access effects on the various transport phenomena.

\section{Unified methodology for non-equilibrium MD simulations across a nanopore}
	We will investigate the flux properties through a CNM nanopore under different stimuli: pressure, chemical or electrical potential difference as illustrated in figure~\ref{fig:system}a.

	\subsection{The nanofluidic periodic model}
\begin{figure}[h!] 
			\begin{small}
				\begin{center}
					\includegraphics{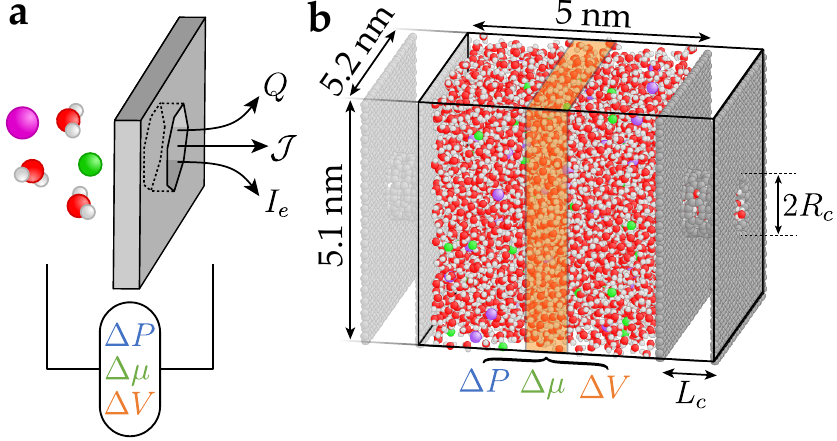}
				\end{center}
				\caption{
					(a) Cartoon of the CNM pore through which different observables like water volume flux $Q$, excess ionic flux $J$ and ionic current $I_e$  induced by either pressure $\Delta P$, chemical $\Delta \mu$ or electrical $\Delta V$ potential drop, will be computed. 
					(b) 3D Periodic rectangular cell used in the non-equilibrium MD with its dimensions in $\unit{nm}$. The carbon membrane formed by two porous graphene sheets encapsulating the CNM pore is placed in the right side. On the left side a liquid slab fills the resting space. The carbon membrane is also repeated with a lighter color on the left side to show the periodicity of the system along the z axis. The orange highlighted area marked the control slab : volume inside which external forces are applied on each species. See figure caption \ref{fig:nanopore} for atom color codes.}
				\label{fig:system}
			\end{small}
		\end{figure}

		The nanofluidic model shown in figure~\ref{fig:system}b is built as follows. The regular CNM nanopore is positioned and aligned with the z direction of space (horizontal axis in the simulation box) and sandwiched between two vertical porous graphene sheets to ensure the sealing. A reservoir of water and ion molecules is placed in contact with the as-formed pierced membrane. The radius of the nanopore can vary between $0.4 \unit{nm}$ and $1.0 \unit{nm}$ (figure~\ref{fig:nanopore}d). 

	\subsection{The mechanical modelling of the flow stimuli}
\begin{figure*} 
			\includegraphics{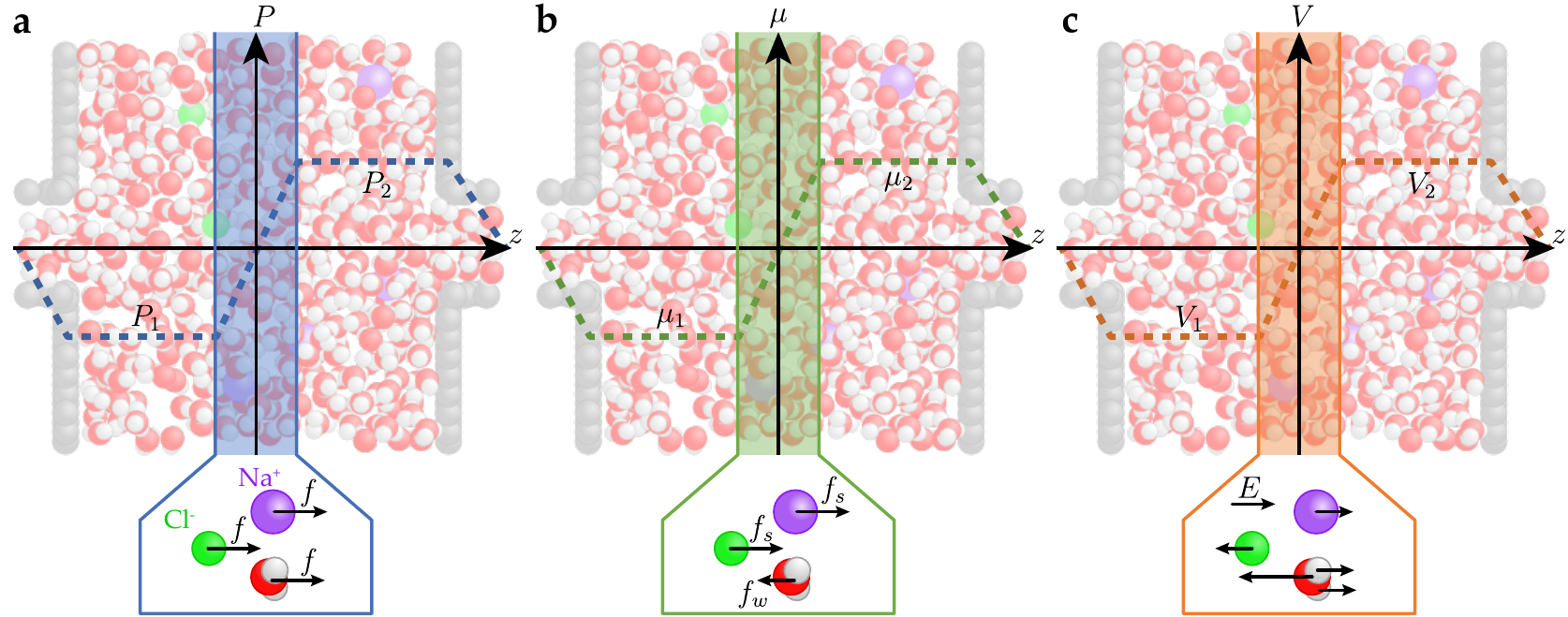}
			\caption{
				Illustration of the non-equilibrium simulation methods for applying a pressure (a), chemical potential (b) and electrical potential (c) drops.
				The colored and semi-transparent areas indicate the control region where the force fields, displayed below each sub-figures, are applied.}
			\label{fig:NEMD}
		\end{figure*}

    We generalize Goldsmith and Martens's approach \cite{goldsmith_martens_2009} to apply a pressure, chemical and electrical potential drop across a nanopore. We define a slab of width $l$ inside the reservoir -- here called the control slab and located in the center of the liquid reservoir--, in which we apply a set of molecular forces to all atoms(see figure~\ref{fig:system}b and \ref{fig:NEMD}). 
    The various thermodynamic drivings are then obtained by choosing specific sets of molecular forces inside the slab.\par

    \par
    
		\emph{Pressure drop $\Delta P$ -} 
		Let us illustrate the spirit of this methodology through the example of pressure-driven flows.

		In order to study flow under a pressure drop $\Delta P$ across a nanoporous membrane, one applies a molecular force $f$ to all atoms in the control slab. This can be understood in simple terms.
		In a translationally invariant system, one can indeed simulate a pressure driven flow by applying a force $f$ along $z$ to all particles (water molecules and ions) in the simulation box. The force balance shows acccordingly that the equivalent pressure drop $\Delta P$ is related to the applied molecular force $f$ according to $\Delta P = \frac{N f}{A}$.  
		Now, in a non-translationally invariant system, {\it e.g.} involving a nanopore across a membrane as considered here, the slab of finite width in the reservoir acts as a 'bulk' system inside the reservoir (see figure~\ref{fig:NEMD}a).
		A pressure drop $\Delta P$ is built across this slab and a simple force balance shows the relation to the force $f$ is the same as above \cite{zhu_schulten_2002, goldsmith_martens_2009, suk_aluru_2010},
		\begin{equation}
			\Delta P = \frac{N f}{A}.
		\end{equation} 
		$N$ is the number of particles {\it inside} the control slab and $A$ is the cross-section area.
		Accordingly an opposite pressure drop is generated across the complementary part of the system, including the nanoporous membrane. This builds a (periodic)  pressure profile across the whole system as sketched in figure \ref{fig:NEMD}a.\par

		This methology can be generalized to any thermodynamic driving force by applying a specific set of mechanical forces to all atoms.\par

		\emph{Chemical potential drop $\Delta \mu$ -} We adapt an approach used to simulate diffuso-osmosis near a surface \cite{yoshida_bocquet_2017} in order to create a chemical potential drop. In the control slab, we apply a force $f_w$ along $z$ to all water molecules and a counterforce $-f_s$ to the ions such that the sum of the external forces is zero (figure~\ref{fig:NEMD}b). Next, we relate the magnitude of the external force to the osmotic pressure difference $\Delta \Pi$ :
		\begin{equation}
			N_w f_w = - N_s f_s = A \Delta \Pi.
		\end{equation}
		$N_s$ and  $N_w$ are the numbers of solute particles and water molecules inside the control slab.
		The force can be related to the concentration difference across the control zone using the van~'t Hoff (or Gibbs-Duhem) relation:
		\begin{equation} \label{eqn:vanthoff}
			\Delta \Pi = \bar{\rho}_s \Delta \mu = k_B T \bar{\rho}_s \Delta \ln \rho_s 
			\underset{\mathrm{dilute}}{\approx} k_B T \Delta \rho_s 
		\end{equation}
		with $\rho_s = \rho_{Na} + \rho_{Cl}$ the concentration of the solute, $\Delta \rho_s$ the concentration drop, $\bar{\rho}_s$ the bulk density and $\Delta \mu$ the chemical potential difference.The corresponding chemical potential profile across the whole system is sketched in figure \ref{fig:NEMD}b.\par

		\emph{Electrical potential drop $\Delta V$ -} To reproduce an electric potential drop $\Delta V$, we propose likewise to use a force-derived approach: an electric field $E = \frac{-\Delta V}{l}$ is applied in the control slab where $l$ is its thickness. If $\Delta V >0$ (as shown in figure~\ref{fig:NEMD}c), the electric field will create an opposite move of the ions and hence a charge imbalance on both sides of the system. Consequently, there will be more $\mathrm{Na^+}$ ions downstream of the control slab and more $\mathrm{Cl^-}$ ions upstream. Thus, one can compare the impact of this control layer to that generated by electrodes on either side of a system that imposes a Nernst potential drop. The corresponding electric potential profile across the whole system is sketched in figure \ref{fig:NEMD}c. We checked that results were independent of the width of the control slab, provided it is larger than the molecular size (say, larger than $0.5\unit{nm}$). Other works apply a homogeneous electric field throughout the cell\cite{sathe_schulten_2011} and we verified that results are similar in terms of fluxes and electric profile. In the non-equilibrium situation under consideration, there are non-zero ionic fluxes, and the electric potential is determined by the conservation laws governing these fluxes. Unlike in equilibrium, there is no formation of an electric double layer, and the resulting electric potential profile is not screened. Just like $\Delta P$ and $\Delta \mu$, we denote $\Delta V$ as the electric potential difference generated by an external thermodynamic force acting on the system. Note that, as in the experiments, these externally applied quantities do not account for the local system's response, which involve the susceptibility of the fluid.\par
		
		Altogether these three methods provide a simple \emph{unified} approach to characterize the flow of water and ions through a nanofluidic membrane under multiple external constraints, keeping a single simulation cell. We hence differ from traditional methods that rely on two separate reservoirs with different thermodynamic conditions, typically different solute concentrations \cite{shen_lueptow_2016, wu_matsuyama_2021}, or adjustable reservoir size \cite{kalra_hummer_2003, raghunathan_aluru_2006}. In short, our approach allows to perform out-of-equilibrium simulations at \emph{steady state} and provides multiple benefits: only one reservoir is simulated, statistics can be accumulated over long times and long time dynamics can be run.\par

	\subsection{MD simulation settings}

		\subsubsection{Classical MD}
			We perform molecular dynamics simulations using a modified version of GROMACS 2021 package.\cite{abraham_lindahl_2015, monet_gromacs_2023}
			Dispersion interactions are modeled using effective Lennard-Jones potentials truncated at $1.2 \unit{nm}$ using a Verlet cutoff scheme. The Coulomb force is treated using a real-space cutoff at $1.2 \unit{nm}$ and particle mesh Ewald summation (pseudo-2D  particle mesh Ewald \cite{york_pedersen_1993}). We use the three-site model SPC/E for water molecules. 
			The Lennard-Jones interaction parameters for NaCl ions are those given by Smith and Dang \cite{smith_dang_1994}. We consider the parameters given by Werder et al. \cite{werder_koumoutsakos_2003} for carbon atoms. The carbon atoms, which constitute the walls and the nanopore, are kept frozen during the simulation.\par
			The thermalization process is done as follows. A first simulation in the canonical ensemble (NVT) of  $1 \unit{ns}$ duration with a time step of $1 \unit{fs}$ is performed. During this run, the system is coupled to a thermostat at $300 \unit{K}$ using velocity rescaling with a stochastic term\cite{bussi_parrinello_2007}. Such initial MD run permits to ensure liquid equilibrium, i.e. the mass density at the center of the reservoir is equal to that of bulk water ($\sim 1 \unit{g/cm^3}$) and that the NaCl ion concentration is equal to $1 \unit{mol/L}$. Under these conditions, there are more than 4000 water molecules and about 60 Na and Cl ions in the system.
		
			\subsubsection{Computing flux averages }
					\begin{figure}[h!] 
					\begin{small}
						\begin{center}
							\includegraphics{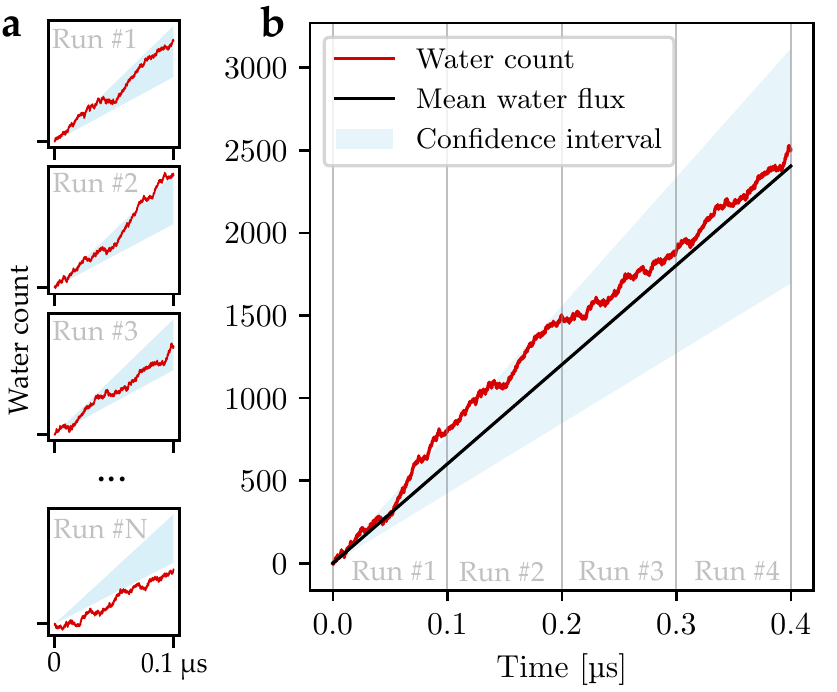}
						\end{center}
						\caption{
							(a) Raw number of water molecules (red curve) that passed through the CNM membrane during simulations for $R_c=0.55\unit{nm}$ (CNM9) and under a pressure stimulus of $10\unit{MPa}$. 
							(b) Concatenation of the few simulations with different initial configurations from which the average flux (black line) and its uncertainty (blue area) are computed.}
						\label{fig:flux}
					\end{small}
				\end{figure}
			Under an external force field, water molecules and ions are flowing through the nanopore. We hence perform a simulation using the canonical set ($300 \unit{K}$) of $1 \unit{ns}$ with a time step of $1 \unit{fs}$. At the end of the simulation, we check that the system has reached a steady state, i.e., that the particle flux has reached a constant value (within statistical fluctuations). Then we perform a long-time simulation ($100 \unit{ns}$) with a time step of $2 \unit{fs}$. According to the desired statistics on the particle flux, we perform several simulations (about 15 runs) in parallel with different initial atomic configurations.\par

			To determine the average particle flux, we simply count the number of particles that passed through the nanopore during the simulations. When several simulations that share the same configuration (apart from the initial atomic distribution) are performed in parallel, the results are concatenated (see figure~\ref{fig:flux}). Then, we compute the average flux of water molecules and Na/Cl ions through the system. The confidence interval is computed from the block average method to eliminate correlations at short times. \cite{flyvbjerg_petersen_1989}

			\subsubsection{Homogeneous fluid approximation model using particle depletion lengths}
					\begin{figure}[h!] 
					\begin{small}
						\begin{center}
							\includegraphics{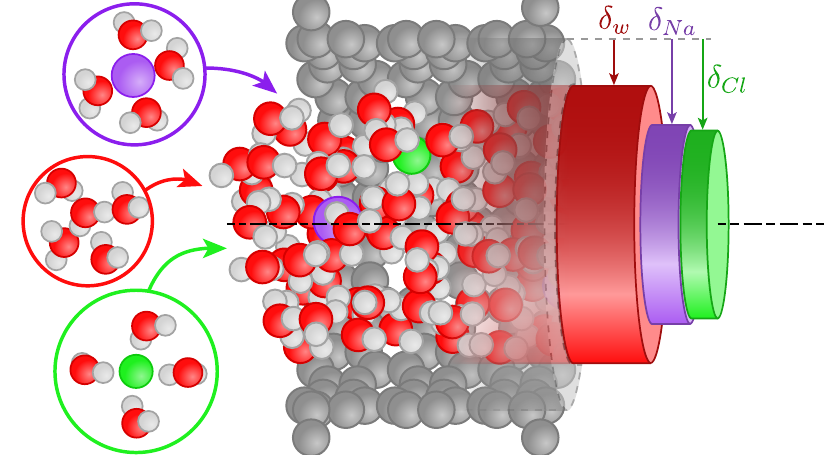}
						\end{center}
						\caption{
							Homogeneous fluid model: (Left) Cartoon of explicit water molecules and the solvated ions that enter a nanopore. (Right) Sketch of the equivalent homogeneous fluid model with corresponding depletion lengths $\delta_i$ with $i=w, Na, Cl$
						}
						\label{fig:model}
					\end{small}
				\end{figure}
			The NEMD simulation results presented below will be analyzed in the light of a continuum model. We model the particle flux $i$ passing through a nanopore of radius $R_c$ by an homogeneous fluid of radius $R_i=R_c-\delta_i$ with density equal to that in the bulk (figure~\ref{fig:model}). $\delta_i$, the depletion length represents the characteristic distance separating the considered species (solvent and solute particles) from the carbon walls. More precisely the depletion length definition \cite{janecek_netz_2007} is adapted for a cylindrical geometry such that 
			\begin{equation}
				\pi R_i^2 = \pi \left(R_c - \delta_i\right)^2 = 2 \pi \int_0^{R_c} {\frac{\rho_i(r)}{\bar{\rho}_i} \, r \, dr}.
			\end{equation}
			$\rho_i(r)$ is the average density of water molecules ($i=w$) or solute particles ($i=Na,Cl$) in the nanopore at a distance $r$ from the pore axis. \par 
			To compute these depletion lengths, we performed a $15 \unit{ns}$ molecular dynamics simulation at equilibrium with a time step of $2 \unit{fs}$ considering a $0.93 \unit{nm}$ radius nanopore (CNM15).  
			As a result, we obtain for water a depletion length $\delta_w=0.20\unit{nm}$ in agreement with the measured values on hydrophobic surfaces such as graphene \cite{janecek_netz_2007, huang_bocquet_2008}. 
			For ionic solutes, the depletion lengths are larger than that of water molecules because ions enter the nanopores 'dressed' with their solvation sphere and we obtain the following values :
			$\delta_{Na} = 0.49\unit{nm}$ and $\delta_{Cl}=0.55\unit{nm}$.
			As illustrated in figure~\ref{fig:model}, there is a difference in the size of the solvation sphere and the depletion length between these ions. The solvation sphere of Na has inward-pointing oxygen atoms, while for Cl ions, it is one of the water's hydrogen atoms that points inward, leading to an expansion of the water shell.
			This model captures well the specific and main effect of nanoporous membranes: the ability to separate ions from water molecules.
			With this model, we can also predict the effect of an evenly distributed surface charge on the nanopore: the depletion length would further depend on the sign of the surface charge attracting (repelling) mobile ions with opposite (same) signs respectively. 
		
\section{The effect of pressure drop $\Delta P$}
	To start we consider the most common stimuli - an applied pressure - onto an electrolyte, since it mimics the principles of a standard  desalination process using Reverse Osmosis. Using the methodology described above, we have performed NEMD simulations under the effect of a pressure drop. The applied pressure should lie within the linearity of the flux response and this constraint leads us to consider a pressure difference of $10 \unit{MPa}$ in the following (see details in appendix~\ref{section:DeltaP-linearity}). \par
\begin{figure}[h!] 
		\begin{small}
			\begin{center}
				\includegraphics{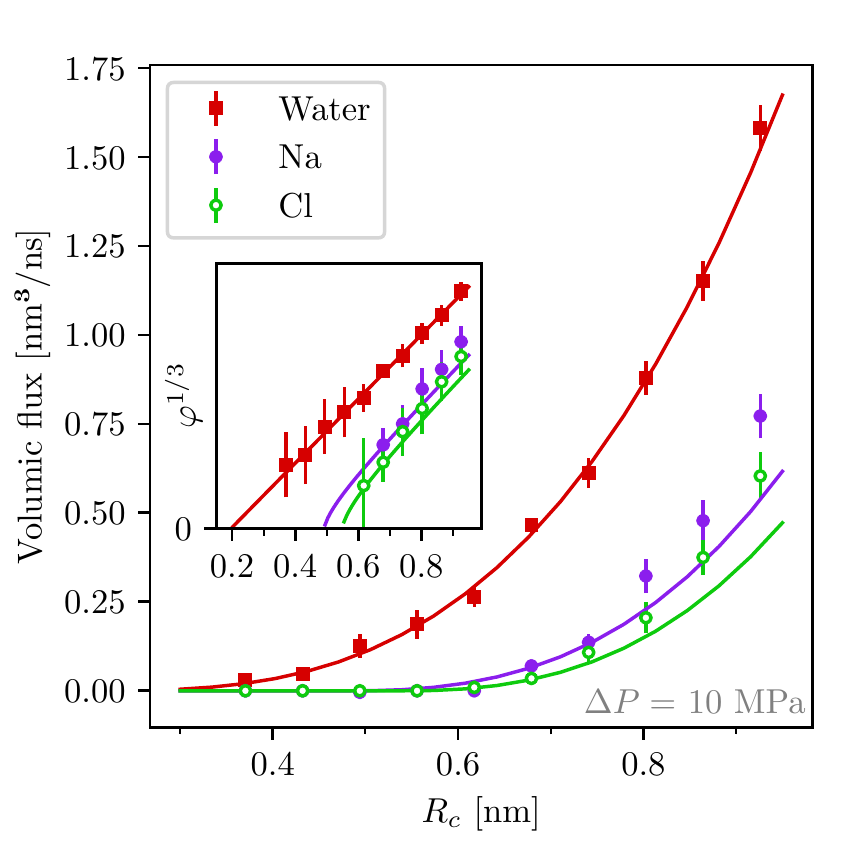}
			\end{center}
			\caption{
				Particle volume flux (water and ions) as a function of pore radius under the effect of a $10 \unit{MPa}$ pressure drop (colored symbols). The continuum model is shown with a corresponding colored solid line. 
				(Inset) the volume flux at power $1/3$ is plotted against the radius and displays linearity.
			}
			\label{fig:DeltaP-flux_v}
		\end{small}
	\end{figure}

	The figure~\ref{fig:DeltaP-flux_v} shows the particle volume flux $\varphi_i$ of particles $i$ as a function of nanopore radii $R_c$. It is defined as the molecular flux $\dot{N}_i$ divided by the volume density in the bulk $\bar{\rho}_i$. The particle volume flux allows to compare the flux of all particles on a single scale. As expected the flux increases with the pore size and exhibits a limit radius ($0.5 \unit{nm}$) below which the ions can not enter. Since CNM membranes consist of pores with radius approximately equal to $0.4 \unit{nm}$ (CNM7), our simulations predict that CNM membranes are impermeable to the salt, in complete agreement with the measurements \cite{yang_golzhauser_2020}.
	Moreover, we can derive the permeability of a nanopore $\mathcal{L}$ such that $\varphi_w = \mathcal{L} /\Delta P$. Thus, for a nanopore $N_t=7$ (CNM7), we obtain $\mathcal{L} \approx 160 \unit{molecules/pore/\mu s/MPa}$. This value is in the range of the experimental measurements; the pore density for a CNM membrane being $0.1-1\unit{pores/nm^2}$, the experimental permeability is around $70-700 \unit{molecules/pore/\mu s/MPa}$ \cite{yang_golzhauser_2018}.\par

	As shown in the inset of figure~\ref{fig:DeltaP-flux_v}, the particle flux scales as $R^3$ and we can reproduce theoretically these results on the basis of entrance effects. For a nanotube of radius $R$ and length $L$, the permeability is indeed the sum of two components \cite{kavokine_bocquet_2021}
	\begin{equation}
		\mathcal{L}^{-1} = \mathcal{L}^{-1}_{\mathrm{entrance}} + \mathcal{L}^{-1}_{\mathrm{friction}},
	\end{equation}
	where
	$\mathcal{L}_{\mathrm{entrance}}$ is the entrance effect according to Sampson's law,
	\begin{equation}
		\mathcal{L}_{\mathrm{entrance}} = \frac{R^3}{3 \eta},
	\end{equation}
	with $\eta$ the viscosity of water,
	and  $\mathcal{L}_{\mathrm{friction}}$ is the friction component of the water on the nanotube walls.
	The latter component can be written as a Hagen-Poiseuille equation with a finite slip condition at the boundaries,
	\begin{equation}
		\mathcal{L}_{\mathrm{friction}} = \frac{\pi R^4}{8 \eta L} \left(1+\frac{4 b}{R}\right)
		\label{friction}
	\end{equation}
	with $b$ the slip length.
	Note that the slip contribution to equation~\ref{friction} actually rewrites as $ {\pi R^3}/{2 \lambda L} $, where $\lambda=\eta/b$ is the water surface friction on the pore's wall. This surface term is dominant over the non-slip part equation~\ref{friction} for very small radius $R$ and its expression does not rely on the validity of the Navier-Stokes equation at the smallest scales.\par
	In order to compare these two components, we need to evaluate the slip length $b$ or friction $\lambda$. Therefore we performed additional MD simulations of water flow through an infinite nanotube. The nanotube is based on the structure of a CNM nanopore shown in figure~\ref{fig:system}b-c where the TPT chains have been extended to infinity along the nanopore axis via periodic boundary conditions. Under the effect of an acceleration field of $100 \unit{nm/ns^2}$, the water molecules reach a terminal velocity. In this situation, the external acceleration field and the friction force compensate each other and one can deduce the friction coefficient from the value of the terminal velocity. We accordingly deduce the slip length for nanotubes with different radii ranging from $0.5\unit{nm}$ to $1.0\unit{nm}$. In all cases, we measure a slip length higher than $20 \unit{nm}$. This value is in agreement with what is conventionally computed in carbon nanotubes \cite{kannam_todd_2017}.
 	As a result, for a CNM nanopore ($R, L\sim 1\unit{nm}$), the inner friction contribution to the permeability is negligible against the entrance contribution: typically, one finds that the impact of the entrance component on the total permeability is at least 100 times higher than the Hagen-Poiseuille component. Therefore, we neglect the friction term.\par
	
	In figure~\ref{fig:DeltaP-flux_v}, the simulated water flux is perfectly reproduced (solid curve in red) considering only the entrance effect with the homogeneous fluid model (figure~\ref{fig:model}):
	\begin{equation} \label{eqn:phi_w_DeltapP}
		\varphi_w = \frac{R_w^3}{3 \eta} \Delta P
	\end{equation}
	where $\eta = 0.853 \unit{mPa \cdot s}$ and $R_w = R_c-\delta_w$ is the effective radius of the nanopore for water ($\delta_w=0.2$nm is the depletion layer for water, measured above). The homogeneous model also works well for ions by considering that the velocity of the solvent particles is equal to that of the solute:
	\begin{equation}
		v_{Na/Cl} = v_w \Rightarrow \varphi_{Na/Cl} = \frac{R_{Na/Cl}^2}{R_w^2} \varphi_{w} = \frac{R_{Na/Cl}^2}{R_w^2} \frac{R_w^3}{3 \eta} \Delta P
		\label{eqn:streamflux}
	\end{equation}
	For ionic concentrations up to $1\unit{mol/L}$ as considered here, the ionic fluxes are found to be linear in the bulk ion concentration, and the corresponding volumetric fluxes and ion velocities in equation~\ref{eqn:phi_w_DeltapP} and \ref{eqn:streamflux} are independent of the ion concentration.\par
	As a final note, we address the potential effect of the deformed shape of the nanopore with non-circular shape and a dissymmetry between the entrance and exit sections of the nanopore. We investigated this geometry effect by additional numerical simulations (see Appendix \ref{section:deformation}) and the general conclusion is that the deviation to the circular shape is not an influential parameter, the crucial parameter being the overall surface area of the entrance of the pore validating the regular tubular model for CNM.\par

\section{The effect of chemical potential drop $\Delta \mu$}
	We now consider transport and flows under chemical potential drop $\Delta \mu$. This thermodynamic driving force is relevant for energy-related applications like "blue energy" arising from the mixing between salted water and fresh water \cite{siria_bocquet_2017}. 

	\begin{figure}[h!] 
		\begin{small}
			\begin{center}
				\includegraphics{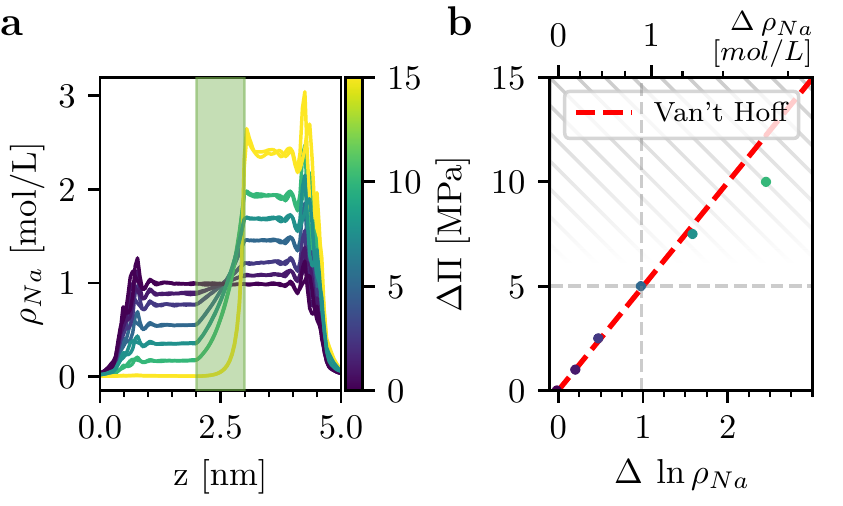}
			\end{center}
			\caption{
				(a) Computation of solute (Na or Cl) concentration (left axis) across the reservoir (light green area) for different osmosis pressure drops (scaled color curves). 
				(b) Osmosis pressure against the log of solution concentration drop. The dashed red line is van~'t Hoff's law.
				Apart from the first few angstroms near the walls, the concentration profile of Cl is very similar to that of Na ions, leading to nearly identical data points on Figure b.
			}
			\label{fig:Deltamu-vanthoff}
		\end{small}
	\end{figure}

	As explained in the methodology section, we apply forces on each species that mimic the osmotic pressure drop $\Delta \Pi$, and which is related to the solute concentration drop across the membrane according to equation~\ref{eqn:vanthoff}. In figure~\ref{fig:Deltamu-vanthoff}a, we plot the concentration profiles of Na and Cl ions along the reservoir as obtained with the simulations. Near the walls, at $z=0\unit{nm}$ and $\unit{5 nm}$, the concentration is zero due to the depletion zone. When approaching the center of the box, we identify two plateaus on both sides of the reservoir which allows defining the concentration drop $\Delta \rho_i = \rho_i(z\in[3.0,4.0 \unit{nm}]) - \rho_i(z\in[1.0,2.0 \unit{nm}])$. Finally, in the control slab at the center of the reservoir -- marked by the green colored area on figure\ref{fig:Deltamu-vanthoff}a--, we identify the transition zone in which we apply the external force field. Figure~\ref{fig:Deltamu-vanthoff}b clearly shows that the concentration drop evolves with $\Delta \Pi$ in accordance with the van~'t Hoff relation (equation~\ref{eqn:vanthoff}) up to $10 \unit{MPa}$. Additionally, we find that the flow response becomes nonlinear above $10 \unit{MPa}$ (see appendix~\ref{section:Deltamu-linearity}).\par

\begin{figure}[h!] 
		\begin{small}
			\begin{center}
				\includegraphics{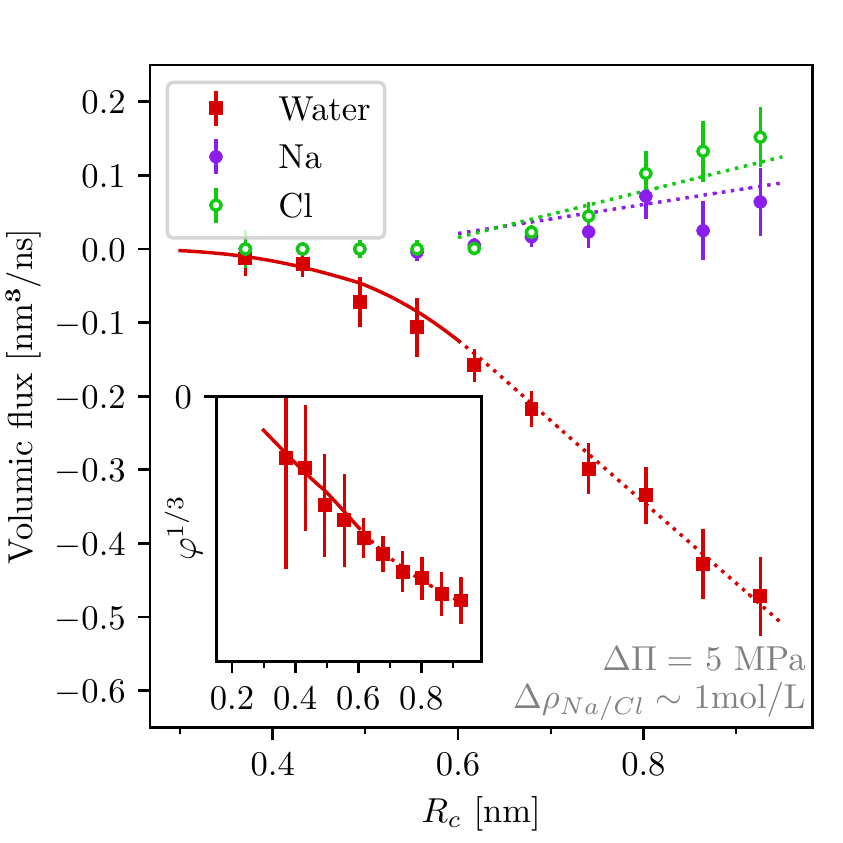}
			\end{center}
			\caption{
				Water and ionic volume flux as function of pore radius under the effect of a $5 \unit{MPa}$ osmosis pressure drop ($\Delta \rho_s \sim 1 \unit{mol/L}$). The model for small pore radius is shown with a solid line. The inset shows the volume flux at power $1/3$. The dotted lines outline the linear dependence of the flux for nanopore radii larger than $0.6 \unit{nm}$.
			}
			\label{fig:Deltamu-flux_v}
		\end{small}
	\end{figure}

	In the following we fix an osmotic pressure of $5 \unit{MPa}$ in order to obtain a solute concentration drop $\Delta \rho_{Na/Cl} \approx 1 \unit{mol/L}$. The NEMD results are displayed in Figure~\ref{fig:Deltamu-flux_v}.\par

	Since ions follow the concentration gradient by diffusion, we observe a positive ion flux that increases with the size of the pores. In contrast, under the effect of osmotic pressure, the water goes in the opposite direction with a more important volume flux with larger pore radii.
	For radii smaller than $0.6 \unit{nm}$, the ion flux is negligible and one can model the water flux in a way similar to the permeability (equation~\ref{eqn:phi_w_DeltapP}),
	\begin{equation}
		\varphi_w = -\frac{R_w^3}{3\eta} \Delta \Pi.
	\end{equation}
	The model is represented by the red solid curve in figure~\ref{fig:Deltamu-flux_v}. For nanopores with larger radii, $R_c>0.6 \unit{nm}$, the flow structure is much more complex. The molecular flux strongly depends on the Debye length $\lambda_D$ and the solute-membrane interaction potential. For the considered system, the Debye length is approximately equal to $0.3 \unit{nm}$, which is smaller than the radius. Assuming that the interaction potential depends mainly on the distance between the solute and the side of the nanopore, Rankin et al. \cite{rankin_huang_2019} found that the flux of water molecules and ions must evolve linearly with the radius of the nanopore. Using the homogeneous fluid model, we should have 
	\begin{equation}
		\varphi_i \propto R_i.
	\end{equation}
	Our simulation results are in qualitative agreement with this relationship with some fluctuations (see dotted line in Figure~\ref{fig:Deltamu-flux_v}).

\section{The effect of electrical potential drop $\Delta V$}
\begin{figure}[h!] 
		\begin{small}
			\begin{center}
				\includegraphics{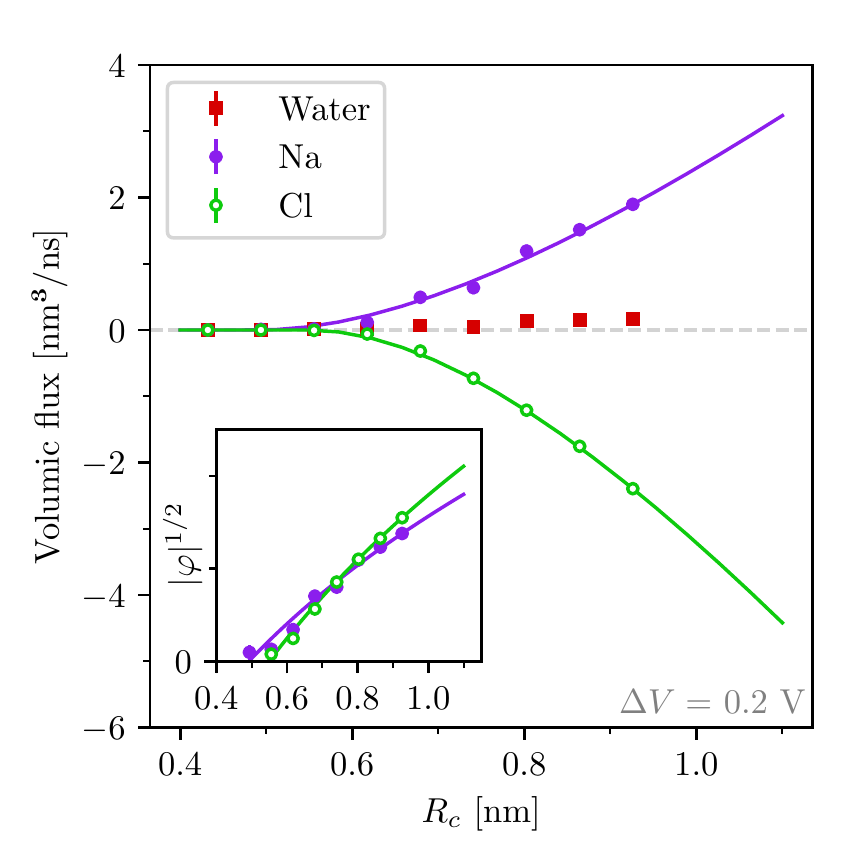}
			\end{center}
			\caption{
				Water and ionic volume flux as a function of pore radius under the effect of a $0.2 \unit{V}$ electrical potential drop. The model is shown with a solid line. The inset shows the square root of the absolute value of volume flux. The volume flow of water under an electrical potential difference is also shown separately in figure~\ref{fig:transport_matrix}d.
			}
			\label{fig:DeltaV-flux_v}
		\end{small}
	\end{figure}

	Finally, we perform a third set of non-equilibrium simulations by imposing an electric potential drop $\Delta V$ across the system.
	Figure~\ref{fig:DeltaV-flux_v} shows the simulation results under the effect of a $0.2 \unit{V}$ potential drop for nanopore radius ranging between $0.4$ to $1.1 \unit{nm}$. We also checked that the system responds linearly for this value of potential drop (see appendix \ref{section:DeltaV-linearity}). 
	
	As expected, the larger the nanopores, the larger the individual ion flux, with opposite signs for $Na^+$ and $Cl^-$ anions. Here we find that the sieving limit, {\it i.e.} the radius beyond which ions can cross the pore, is $0.6 \unit{nm}$,  similar to that in the case of pressure driving force. Furthermore, the water volume flow is found to remain low compared to the solute volume flow. This is expected because the molecules are neutral and the ionic fluxes that can drive water molecules, through the solvation sphere, compensate each other (see the discussion on the streaming current below). 

	To analyze these results we define the conductance $G$ as $I_e = G \Delta V$, with $I_e$ the electric current generated by the circulation of ions through the system. Similarly to the permeability, there are two components to the conductance \cite{kavokine_bocquet_2021} 
	\begin{equation}
		G^{-1} = G^{-1}_{\mathrm{entrance}} + G^{-1}_{\mathrm{volume}}.
	\end{equation}
	The first term involving $G_{\mathrm{entrance}}$, is the entrance effect.
	\begin{equation}
		G_{\mathrm{entrance}} = \sigma_B \alpha R 
	\end{equation}
	with $\sigma_B$ the conductivity and $\alpha$ a constant approximately equal to $2$. We remind that the conductivity can be related to the mobilities of Na and Cl ions ($\mu_{Na}$ and $\mu_{Cl}$) by the following relation: $\sigma_B = e^2 (\bar{\rho}_{Na} \mu_{Na} +  \bar{\rho}_{Cl} \mu_{Cl})$. 

	The second term $G_{\mathrm{volume}}$ is the conventional bulk conductance. It is the conductance of a tube of radius $R$, length $L$ and conductivity $\sigma_B$,
		\begin{equation}
			G_{\mathrm{volume}} = \sigma_B \frac{\pi R^2}{L}.
		\end{equation}

	On the figure~\ref{fig:DeltaV-flux_v}, we find both components: a first quadratic regime for small radii dominated by the bulk conductance then a second linear regime described by the entrance effects.
	We can measure quantitatively this agreement by computing the partial components of Na and Cl ions in the conductance $G_{Na/Cl}$ and their mobility $\mu_{Na/Cl}$. Indeed, assuming that the ionic volume flux is 
	\begin{equation} \label{eqn:mobility}
		\varphi_i = \mu_i \bar{\rho}_{i} z_i \left(\frac{L}{\pi R_i^2} + \frac{1}{\alpha R_i}\right)^{-1} \Delta V
	\end{equation} 
	with $R_i = R_c-\delta_i$ and $z_i$ the ionic valency, we can determine the evolution of the mobility as a function of the nanopore size. \par

	\begin{figure}[h!] 
		\begin{small}
			\begin{center}
				\includegraphics{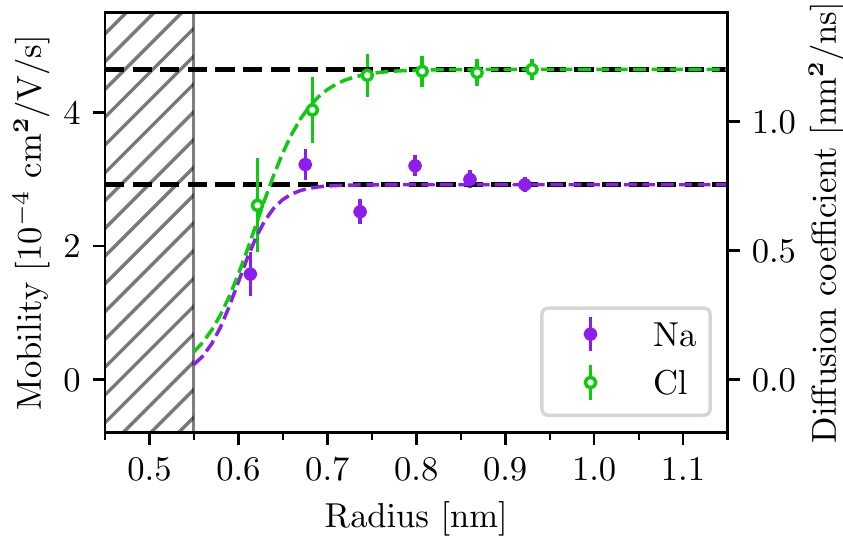}
			\end{center}
			\caption{
				Ionic mobility (left axis) and associated diffusion coefficient (right axis) as a function of the nanopore radius. The hatched area covers the region where the mobility is not defined since the ionic flux is null.}
			\label{fig:mobility}
		\end{small}
	\end{figure}

	Figure~\ref{fig:mobility} shows the evolution of the ion mobility $\mu$ (or equivalently the ion diffusion coefficient $D = \mu k_B T$) as a function of $R_c$ computed from the simulated ionic volume flux compared to the prediction in equation~\ref{eqn:mobility}. For radii smaller than $0.5 \unit{nm}$, the mobility is not defined because there is no ionic flux. Between $0.5$ and $0.7 \unit{nm}$, the mobility increases progressively until reaching a threshold value. Beyond that, the mobility (diffusion coefficient) remains constant with $\mu_{Na} = 2.9\;10^{-4} \unit{cm^2/V/s}$ ($D_{Na} = 0.8 \unit{nm^2/ns}$) and $\mu_{Cl} = 4.4\;10^{-4} \unit{cm^2/V/s}$ ($D_{Cl} = 1.3 \unit{nm^2/ns}$).
	It should be noted that these latter values were computed for a salt concentration fixed at $1 \unit{mol/L}$. Yet it is well known that the diffusion coefficient of NaCl decreases significantly with the concentration \cite{robinson_stokes_1959}. So we have performed additional MD simulations to estimate the diffusion coefficient at infinite dilution. We find $D^0_{Na} = 1.31 \unit{nm^2/ns}$ and $D^0_{Cl} = 1.77 \unit{nm^2/ns}$, which are closer to the experimental data \cite{lide_lide_2008} measured at infinite dilution, i.e., $D^0_{Na} = 1.33 \unit{nm^2/ns}$ and $D^0_{Cl} = 2.03 \unit{nm^2/nm}$.
	
	Gathering all results, one can compare the ionic fluxes to the predictions obtained from equation~\ref{eqn:mobility}, with mobility values fixed to the computed threshold values (see Figure~\ref{fig:mobility}). As shown in figure~\ref{fig:DeltaV-flux_v}, the model (solid color curve) is in very good agreement with the simulation data (color symbols, figure~\ref{fig:DeltaV-flux_v}).
	
\section{Transport matrix}

	\begin{figure*} 
		\begin{small}
			\includegraphics{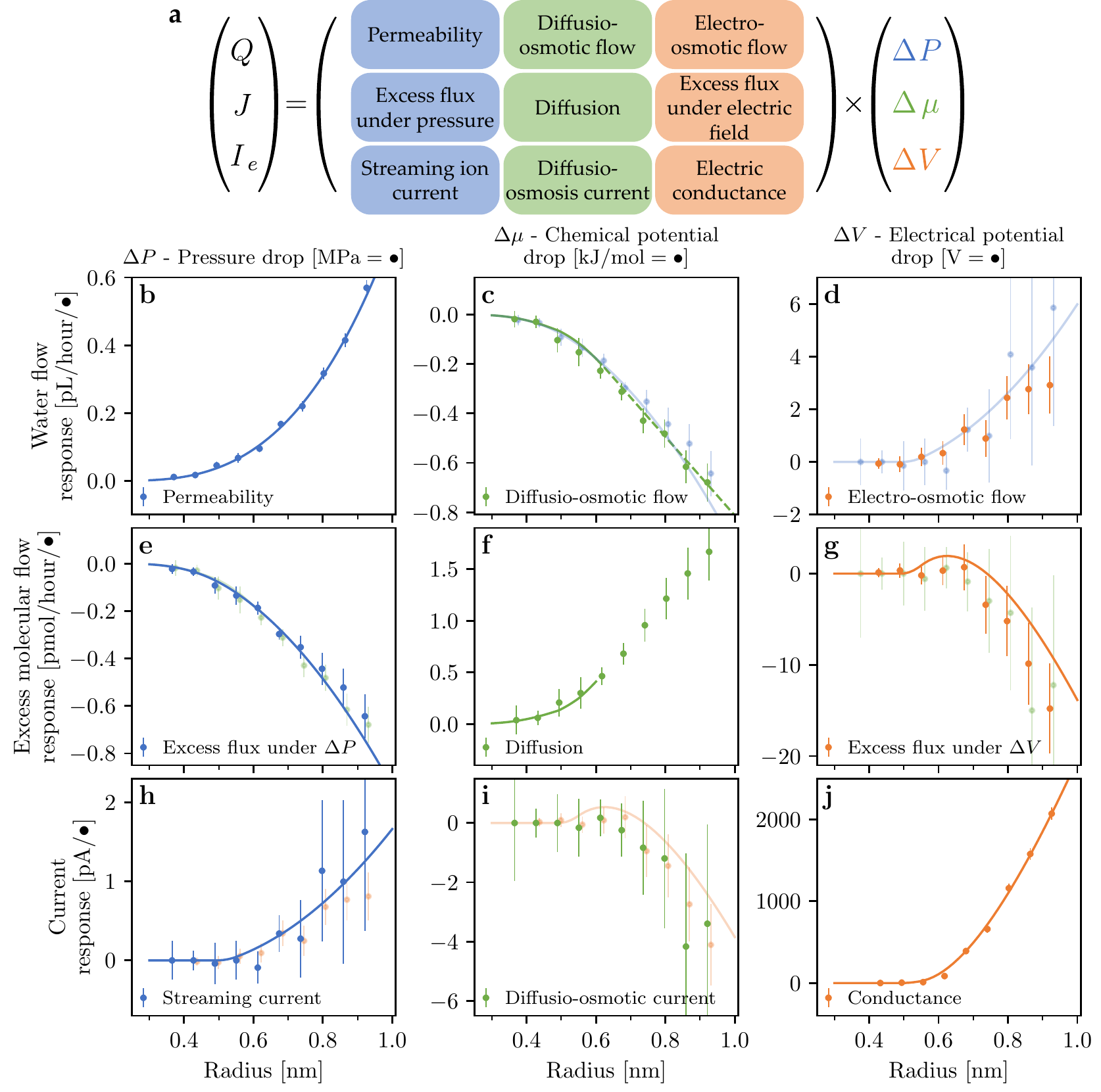}
			\caption{
				(a) Explicit transport matrix.
				(b-j) Response coefficients of water flow, excess ionic flow and electrical current under the effect of pressure, chemical potential and electrical potential drop as a function of nanopore radius. Scatters are related to NEMD and solid lines to theoretical model. On the off-diagonal figures, we show the system response and the reciprocal (or symmetric) response with lighter colors.
			}
			\label{fig:transport_matrix}
		\end{small}
	\end{figure*}
 	So far, we considered the flows of each species (ions and water) separately under the effect of thermodynamic forces. More generally, it is relevant to quantify the complete transport matrix relating fluxes to thermodynamic forces. In linear response, the transport matrix $\mathbb{L}$ is accordingly defined as:
	\begin{equation}
		\begin{pmatrix}
			Q\\J\\I_e
		\end{pmatrix} =
		\mathbb{L} 
		\times
		\begin{pmatrix}
			\Delta P \\ \Delta \mu \\ \Delta V
		\end{pmatrix}.
	\end{equation} 
	The various fluxes which enter the transport matrix are:
	\begin{itemize}
		\item the volume flux of water 
			\begin{equation}
				Q=\varphi_w.
			\end{equation}
		\item the excess ionic molecular flux (in excess to the  convective transport by water)
			\begin{equation}
				J=\bar{\rho}_{Na} \left( \varphi_{Na} - \varphi_w \right) + \bar{\rho}_{Cl} \left( \varphi_{Cl} - \varphi_w \right). 
				\label{defJ}
			\end{equation}
		\item the ionic current 
			\begin{equation}
				I_e = e \bar{\rho}_{Na} \varphi_{Na} - e \bar{\rho}_{Cl} \varphi_{Cl}.
				\label{current}
			\end{equation}
	\end{itemize}
	Due to the Onsager principle, the transport matrix is symmetric and definite positive \cite{onsager_onsager_1931, groot_mazur_1969}. Each term of the matrix corresponds to a specific transport process (see figure~\ref{fig:transport_matrix}a).\par

	Figures~\ref{fig:transport_matrix}b-j show the simulation results for all terms of the matrix, here plotted as a function of the pore radius. On the diagonal, we find respectively the permeability (panel b), the ionic diffusion (panel f) and the ionic conductance (panel j). The off-diagonal terms correspond to cross-terms in the transport: electro-osmosis, streaming currents, diffusio-osmotic flows, diffusio-osmotic currents, etc. These mobilities contain a wealth of information on the molecular mechanism at play and we shortly discuss the behaviors.\par
	
	First, one can check the symmetry of the transport matrix: in the off-diagonal panels of \ref{fig:transport_matrix}b-j, we compare the mobilities to their symmetric counterpart (with lighter curve colors), showing a very good agreement. For example, the excess ionic flux coefficient generated under a pressure drop (panel e) is found to be remarkably equal to the diffusio-osmotic coefficient (panel c), demonstrating the reliability of our simulations. Similarly, the results of the simulations in terms of streaming current (panel h) and, respectively, electro-osmotic flow (panel d) are also very close, as well as the excess ionic flux under a voltage (panel g) and the diffusio-osmotic current under a chemical gradient (panel i).\par
	
	All coefficients are compared to predictions based on continuum modeling described in the previous sections. From the previous predictions for the water, $Na^+$ and $Cl^-$ fluxes under the various thermodynamic drivings, one can calculate the excess ionic current $J$ and $I_e$ as a function of the pore size.\par
	
	The appendix~\ref{section:TM-formula} gathers the expressions of the individual transport matrix coefficient (equations~C1-7). For example, the prediction for the excess flux under a voltage drop (panel g) (equal to the diffusio-osmotic current under a chemical drop, panel i) is obtained by using equations (\ref{eqn:mobility}) and (\ref{defJ}), with the simplifying hypothesis of vanishing water flux in this condition (see equation~\ref{eqn:L_J_DV}). This leads to the solid curve shown in panels g and i, which are in good qualitative and quantitative agreement with the simulation data.
	A similar prediction can be obtained for the streaming currents using the prediction for the $Na/Cl$ fluxes as a function of pressure drop, in equation (\ref{eqn:streamflux}), combined with the definition of the ionic current in equation (\ref{current}), as provided explicitly by equation~\ref{eqn:L_Ie_DP}. \par

	In most cases, one can easily interpret the signs of the mobilities, {\it e.g.} the negative diffusio-osmotic flow, by realizing that the water molecules enter the nanopores more easily than the solute. This reflects the selectivity of the membrane that we simply captured by considering a lower depletion length for water than for ions.
	However, some mobility coefficients are more subtle to interpret physically. For example, the non-vanishing diffusio-osmotic current (panel i) -- and its symmetric in panel g -- is somewhat	unexpected for a neutral nanopore. Usually, such diffusio-osmotic currents under chemical gradients are interpreted in terms of a large surface charge on the pore surface \cite{siria_bocquet_2013,siria_bocquet_2017}. Here a diffusio-osmotic ionic current can be generated under a salt gradient, in spite of the pore being neutral.
	This can be explained by the difference in the interfacial ion profile for the sodium and chloride ions, as inferred from the different values of the depletion layer for the ions $\delta_{Na^+}<\delta_{Cl^-}$. We discuss below some implications of this result.\par
	
\section{Discussion}

	To conclude, we have introduced a molecular dynamics methodology that allows computing the non-equilibrium transport of water and salt through nanoporous membranes.
	The fluid transport can be investigated under different classical driving forces, like pressure, concentration and voltage drops. Our approach is based on a unified mechanical approach, mimicking the thermodynamic drivings via a different set of forces applied on the atoms inside a controlled slab of the liquid reservoir.
	The full transport matrix can be calculated, giving hints on the properties and performance of specific membranes under the different drivings.\par
	
	In the present work, we have applied this methodology to study the properties of carbon nanoporous membranes (CNM), with a focus on a recently synthesized type of CNMs \cite{yang_golzhauser_2018}. These membranes were reported to exhibit outstanding performances for filtration and desalination, with excellent selectivity and considerable permeabilities. Our numerical results for these quantities (selectivity and permeability) are quantitatively consistent with the experimentally reported results. This assesses the considerable potential of such new types of carbon membranes for water treatment.\par
	 
	 	\begin{figure}[h!] 
		\begin{small}
			\begin{center}
				\includegraphics{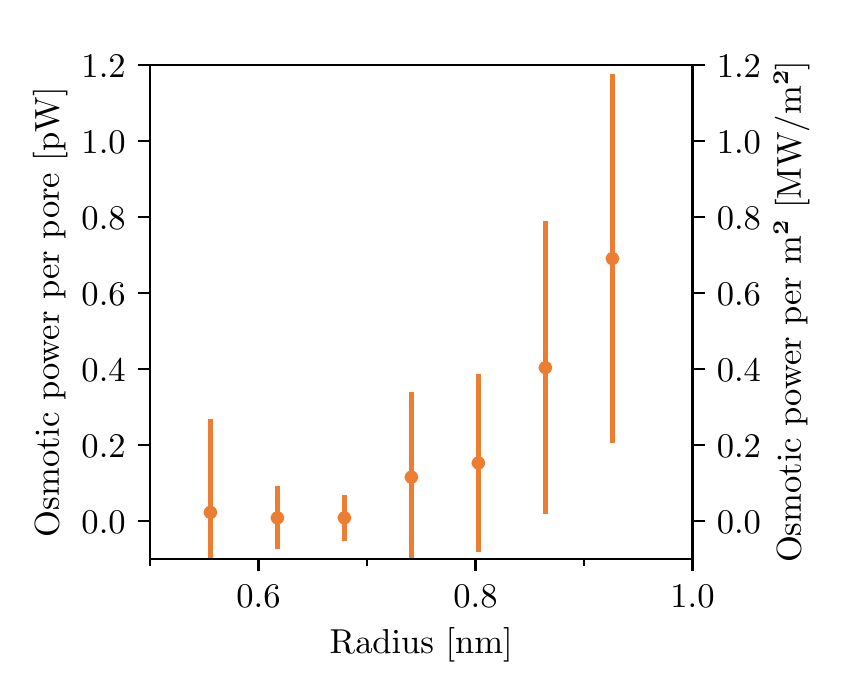}
			\end{center}
			\caption{
				Electric power harvested across a single nanopore in a CNM membrane, under a salinity gradient of 0.015M/0.6M (left axis) and power per square meter (right axis) as a function of the nanopore radius. We use the experimental value for the pore density of $\phi_s\sim 10^{14} \unit{pores/cm^2}$ to extrapolate the power per unit surface. }
			\label{fig:power}
		\end{small}
	\end{figure}

	Now, beyond their performance in terms of desalination, the detailed knowledge of the complete transport matrix provide some hints on the performances of CNM for other applications at the water-energy nexus. In particular, we can make an estimate here of their potential for osmotic power, {\it i.e.} the energy harvested from the mixing of sea and fresh water \cite{siria_bocquet_2013,siria_bocquet_2017}. As shown above, an ionic current $I_e$ is generated across the CNM under a salinity gradient (or chemical potential drop $\Delta \mu$), according to $I_e = {L_{I_e, \Delta \mu}}\times k_BT \Delta \log \rho_s$ (assuming an ideal solution expression for $\Delta \mu$). As commented above, this result is quite counter-intuitive, since the membrane is not charged, and it results from the slight transport asymmetry between sodium and chloride across the carbon nanopores. It can be shown accordingly that the
	maximum electric power which can be harvested takes the expression \cite{siria_bocquet_2017}:
	\begin{equation}
	{\cal P}_{\rm pore}={I_e^2\over 4 G}={ {L_{I_e, \Delta \mu}}^2\over 4 G}\times \left[ k_BT \Delta \log \rho_s\right]^2
	\end{equation}	 
	per nanopore. Extrapolating to a pore density $\phi_s$ of nanopores on the CNM membranes, with $\phi_s\sim 10^{13}-10^{14} \unit{pores/cm^2}$\cite{yang_golzhauser_2018}, one has therefore ${\cal P}_{\rm S}=\phi_s{ {L_{I_e, \Delta \mu}}^2\over 4 G}\times \left[ k_BT \Delta \log \rho_s\right]^2$. 
	We plot in figure \ref{fig:power} the corresponding power -- per pore or per square meter -- for a salt concentration gradient of 0.015M/0.6M across the CNM membrane (sea/river water conditions). The extrapolated osmotic power is found to depend on the pore radius and reaches considerable values reaching MW per square meter, in the same range as the reported value in single nanopores drilled in MoS2 membrane in Ref.~\cite{feng2016single}. 
	Of course, in a large-scale osmotic stack, this performance will be reduced by a number of factors, in particular the various electric resistances in the equivalent circuit, {\it e.g.}, the electric resistances in the low-salinity reservoir, and concentration polarization effects. CNM membranes might also suffer from the usual drawbacks of nanoporous membranes, such as clogging effects. However, they  exhibit outstanding performances in terms of ionic transport, with furthermore some scalability of the fabrication process. This points to the potential of novel nanomaterials, exploiting nanofluidic transport, as new avenues for osmotic energy harvesting.

\begin{acknowledgments}
	We acknowledge the European FET “ITS-THIN” project (N°899528) for funding.
\end{acknowledgments}
	
\section*{Data Availability Statement}
	The data that support the findings of this study are available from the corresponding author upon reasonable request.

\clearpage
\appendix

\section{Linearity limit}
	One could question the linearity of the response of the system under an external perturbation. In the following, we look at the threshold of all stimuli values beyond which the system responds non-linearly.
	\subsection{Pressure drop $\Delta P$} \label{section:DeltaP-linearity}
\begin{figure}[h!] 
			\begin{small}
				\begin{center}
					\includegraphics{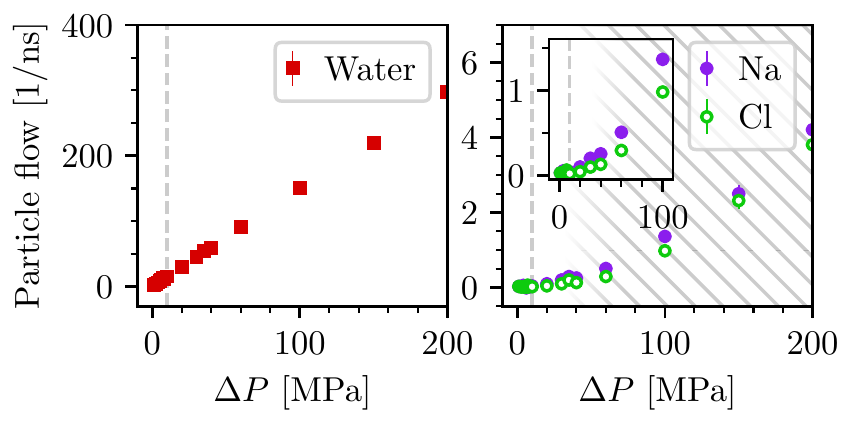}
				\end{center}
				\caption{
					Particle flux of water (a) and ions (b) as a function of osmosis pressure drop. The hatched area covers the non-linear region.
				}
				\label{fig:DeltaP-linearity}
			\end{small}
		\end{figure}

		Figure~\ref{fig:DeltaP-linearity} shows the molecular flux as a function of the pressure drop through the system. We notice that the flux of water molecules responds linearly under a pressure drop even beyond very important values ($>150 \unit{MPa}$). This result is already well known and justifies the use of very high pressure difference in literature. Typically, many molecular dynamics simulations of pure water flow through membranes are performed with pressure differences $\Delta P = 200-300 \unit{MPa}$ to maximize the flux and thus the statistics of the simulations \cite{zhu_schulten_2002,suk_aluru_2010,goldsmith_martens_2009,ritos_reese_2014}.\par

		The situation is different for the ions. On figure~\ref{fig:DeltaP-linearity}b, we see that the ion flux becomes non-linear for much smaller pressure differences. The pressure drop must be less than $50 \unit{MPa}$ to find a linear behavior of the ionic flux as a function of $\Delta P$. Therefore, a pressure difference of $10 \unit{MPa}$ across the system is applied to ensure the linear behavior of the system response. With this value, we approach the pressures typically used in desalination plants.\par

	\subsection{Chemical potential drop $\Delta \mu$} \label{section:Deltamu-linearity}
\begin{figure}[h!] 
			\begin{small}
				\begin{center}
					\includegraphics{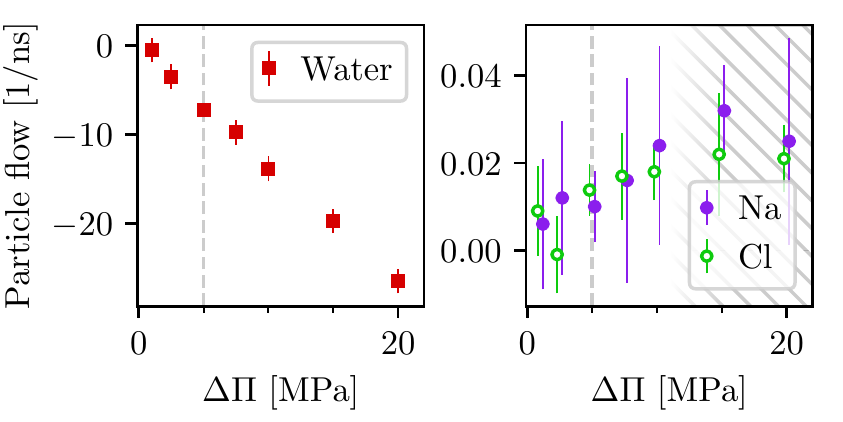}
				\end{center}
				\caption{
					Particle flux of water and ions as a function of osmosis pressure drop. The hatched area covers the non-linear region. 	
				}
				\label{fig:Deltamu-linearity}
			\end{small}
		\end{figure}

		Figure~\ref{fig:Deltamu-linearity} shows that the flux of water and ion molecules scale linearly with osmotic pressure $\Delta \Pi$ up to a pressure of $10 \unit{MPa}$. Beyond this value, the ionic flux seems to saturate. 

	\subsection{Electrical potential drop $\Delta V$} \label{section:DeltaV-linearity}
\begin{figure}[h!] 
			\begin{small}
				\begin{center}
					\includegraphics{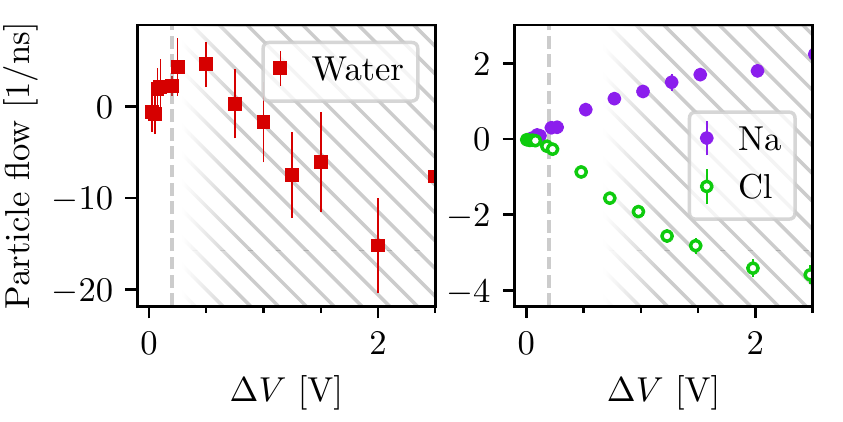}
				\end{center}
				\caption{
					Water and ion molecular flux as a function of electrical potential drop. The hatched area covers the non-linear region.
				}
				\label{fig:DeltaV-linearity}
			\end{small}
		\end{figure}

		 Figure~\ref{fig:DeltaV-linearity} shows that the fluxes of water molecules and ions scale non-linearly beyond an electric potential difference of $0.2 \unit{V}$. Thus, we choose this value to maximize the statistics and keep the linear regime. The simulation box has a depth $\sim 5 \unit{nm}$, a potential difference of $0.2 \unit{V}$ corresponds to an electric field of $\sim 0.04 \unit{V/nm}$. 

\section{Pore shape}\label{section:deformation}
	\begin{figure*}[ht!] 
		\begin{small}
			\begin{center}
				\includegraphics{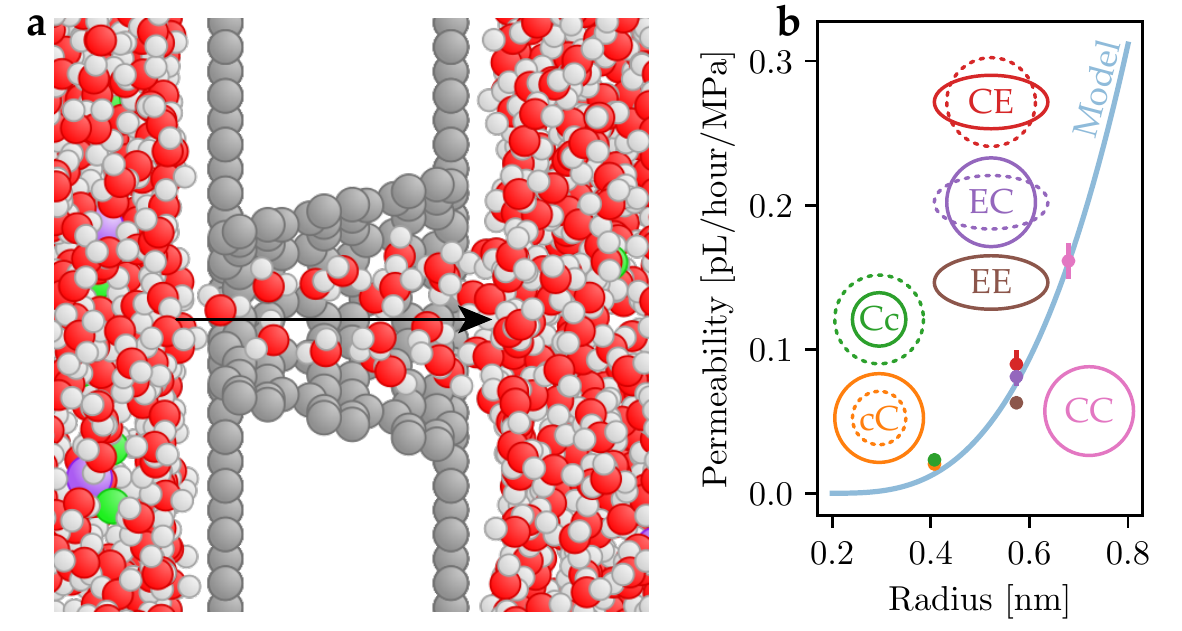}
			\end{center}
			\caption{
				(a) Cross-section view of the cC geometry. The foreground carbon atoms are hidden to reveal the flow of water through the nanopore.
				(b) Measured permeability of the deformed geometries (colored points) compared to the homogeneous fluid model for the permeability (blue curve). The geometries are depicted by drawings such that the colored solid curve, respectively dotted curve, corresponds to the geometry of the exit (right side on the view (a)), respectively the entrance (left side), of the nanopore.}
			\label{fig:deformation}
		\end{small}
	\end{figure*}
	In the following, we question the effect of possible deviations from the perfect circular nanopore and how these geometric changes do affect the results.
	Indeed electron microscopy observations \cite{yang_golzhauser_2018} show deformed nanopores compared to our perfectly circular model. We wish to quantify the effect of these deformations on molecular transport. We start with an initial nanopore with a radius equal to $0.68 \unit{nm}$ on both ends (notation CC) and we deform it. The inlet or outlet radius can be shrunk to a radius equal to $0.41 \unit{nm}$ (notation "c") or ovalized at a constant perimeter (half-width $a=0.87 \unit{nm}$, half-height $b=0.41 \unit{nm}$, notation "E"). We apply a linear transformation on the intermediate carbon atoms in order to ensure the continuity between the inlet and outlet sections. Thus we study 5 deformed systems, cC (shown in figure~\ref{fig:deformation}a), Cc, EE, EC, CE and the corresponding shapes are sketched below their labels in figure~\ref{fig:deformation}b.
	In Figure~\ref{fig:deformation}b, we plot the permeability of the deformed systems as a function of the equivalent radius $R$ corresponding to the minimum cross-section area accessible by the water flow, given the depletion length $\delta_w$, 
	\begin{equation}
		\pi (R-\delta_w)^2 = \underset{j=inlet,\,outlet}{\mathrm{min}} \left \{ \pi (a_j-\delta_w) (b_j-\delta_w)  \right \}.
	\end{equation}
	$a_j$ and $b_j$ are respectively the half-width and half-height of the inlet ($j=inlet$) or outlet ($j=outlet$) section. By defining the equivalent radius, one can directly compare the simulated permeabilities with the one computed from the homogeneous fluid model (equation~\ref{eqn:phi_w_DeltapP}). Clearly, despite the large deformations induced on the nanopore, the simulated points are very close to the theoretical curve. Thus, the permeability of a deformed nanopore corresponds to that of a regular nanopore whose cross-section area is equal to the smallest cross-section area of the deformed nanopore. This result underlines the relevance of our model to provide quantitative interpretations of the molecular flows across real CNM membranes.

\section{Equation of transport matrix components} \label{section:TM-formula}

\subsection{Pressure drop $\Delta P$}

Permeability (figure \ref{fig:transport_matrix}b): 
\begin{equation}\label{eqn:L_Q_DP}
	L_{Q, \Delta P} = \frac{R_w^3}{3 \eta}
\end{equation}

Excess flux under $\Delta P$(figure \ref{fig:transport_matrix}e): 
\begin{equation}\label{eqn:L_J_DP}
	L_{J, \Delta P} = \frac{R_w^3}{3 \eta} 
		\sum_i {\bar{\rho}_i \left(\frac{R_i^2}{R_w^2} -1 \right)}
\end{equation}

Streaming current (figure \ref{fig:transport_matrix}h): 
\begin{equation}\label{eqn:L_Ie_DP}
	L_{I_e, \Delta P} = e \frac{R_w^3}{3 \eta} 
	 \sum_i {\bar{\rho}_i z_i \frac{R_i^2}{R_w^2}}
   \end{equation}

\subsection{Chemical potential drop $\Delta \mu$}

Diffusio-osmotic flow (figure \ref{fig:transport_matrix}c): 
\begin{equation}\label{eqn:L_Q_Dmu}
	L_{Q, \Delta \mu} = \left\{
			\begin{array}{ll}
					- \bar{\rho}_s \frac{R_w^3}{3 \eta} & \mbox{if } R<\delta_{Na/Cl} \\
					\propto -R & \mbox{if } R \gg \delta_{Na/Cl}
			\end{array}
	\right.
\end{equation}

Diffusion (figure \ref{fig:transport_matrix}f): 
\begin{equation}\label{eqn:L_J_Dmu}
	L_{J, \Delta \mu} = \frac{R_w^3}{3 \eta} 
		\sum_i{
			\frac{\bar{\rho}_i}{\bar{\rho}_s}
			\left(\frac{R_i^2}{R_w^2} -1 \right)
		} \mbox{ if } R<\delta_{Na/Cl}
\end{equation}

\subsection{Electrical potential drop $\Delta V$}

Excess flux under $\Delta V$ (figure \ref{fig:transport_matrix}g): 
\begin{equation}\label{eqn:L_J_DV}
	L_{J,\Delta V} = \sum_i{\bar{\rho}_i \mu_i z_i
		\left(\frac{L}{\pi R_i^2} + \frac{1}{\alpha R_i}\right)^{-1}
	}
   \end{equation}

Conductance (figure \ref{fig:transport_matrix}j): 
\begin{equation}\label{eqn:L_Ie_DV}
	L_{I_e,\Delta V} = e\sum_i{\bar{\rho}_i \mu_i z_i^2 
		\left(\frac{L}{\pi R_i^2} + \frac{1}{\alpha R_i}\right)^{-1}
	}
   \end{equation}

$R_j= R - \delta_j$ is the nanopore radius from which the depletion length $\delta_j$ has been subtracted. The sum over $i$ ($\sum_i$) includes all electrolytes (for example $i=Na$ and $Cl$). $\bar{\rho}_i$, $\mu_i$ and $z_i$ are respectively the bulk density, the mobility and the valency of the electrolyte $i$. $\bar{\rho}_s = \sum_i{\bar{\rho}_i}$ is the total bulk density of electrolytes. $\eta$ is the water viscosity and $e$ the elementary charge. $\alpha$ is a geometric constant approximately equal to $2$.
\bibliography{main}
\end{document}